\documentclass[twocolumn,aps,showpacs]{revtex4}
\usepackage{epsfig,color,ulem}

\begin{document}
\title{High $p_T$ resonances as a possibility to explore hot and dense nuclear matter}

\author{S.Vogel ${}^{1}$, J. Aichelin${}^{2}$ and M. Bleicher ${}^{1}$}

\address{${}^{1}$Institut f\"ur Theoretische Physik, J.W. Goethe Universit\"at, \\
Max-von-Laue-Stra\ss{}e 1,\\
60438 Frankfurt am Main, Germany}

\address{${}^{2}$SUBATECH,
Laboratoire de Physique Subatomique et des Technologies Associ\'ees \\
University of Nantes - IN2P3/CNRS - Ecole des Mines de Nantes \\
4 rue Alfred Kastler, F-44072 Nantes Cedex 03, France}

\begin{abstract}
One of the fundamental objectives of experiments with ultrarelativistic heavy ions is to explore strongly interacting matter at high density and high temperature. In this investigation we study in particular the information which can be obtained by analyzing baryonic and mesonic resonances. The decay products of these resonances carry information on the resonances properties at the space time point of their decay. We especially investigate the percentage of reconstructable resonances as a function of density for heavy ion collisions in the energy range between $E_{lab}$ = 30~AGeV and $\sqrt{s}$ = 200~AGeV, the energy domain between the future FAIR facility and the present RHIC collider.

\vspace{.6cm}
\end{abstract}


\maketitle

The experimental analysis of heavy ion reactions using resonances has been applied for several years from low energy \cite{Agakichiev:2006tg,Lopez:2007zz} through intermediate \cite{Afanasev:2001qj,Adamova:2002kf} to high energy heavy ion collisions \cite{Adams:2006yu,Abelev:2008yz,Fachini:2008zz}. In general one
distinguishes between leptonic and hadronic decay channels. While the hadronic  decay channels have the advantage of larger branching ratios, the leptonic decays have the advantage that the decay particles do not undergo final state interactions.
Thus it is worthwhile to work out the differences and the advantages and disadvantages of the two approaches which will be discussed in the following.

The present Relativistic Heavy Ion Collider (RHIC) at Brookhaven and the
upcoming Facility for Antiproton and Ion Research (FAIR, for a recent status on the project we refer to \cite{Henning:2008zz}) provide an excellent research environment for probing resonances in matter.
At the RHIC experiments it has been observed \cite{Markert:2007qg} that less resonances are measured
than expected from statistical model calculations \cite{Andronic:2002pj}. Stable hadrons however follow the prediction of this model. This suggests the conclusion
that after chemical freeze-out, when the chemical composition of the final state is determined, hadrons still undergo collisions and therefore some of
the resonances cannot be identified by the invariant mass of the decay products.

At FAIR the leptonic as well as the hadronic decay channel can be explored.
 While the leptonic channel is usually regarded as the 'cleaner' channel recent calculations \cite{Vogel:2007yu} have shown that the dilepton channel might not probe the dense phase as it was expected before. \\
In light of this new development it is worthwhile to evaluate the density-profile and the space-time-evolution of resonances which can be reconstructed in the hadronic decay channels. Although those channels suffer from the drawback of final state interaction of the decay products, their large branching ratios might make them better suited for the investigation of the high density phase of heavy ion collisions compared to leptonic decay channels.

For our calculations we utilize the UrQMD(v2.3) model, a non-equilibrium transport approach, which relies on the covariant Boltzmann equation. All cross sections are calculated by the
principle of detailed balance and the additive quark model or are fitted to available data.
UrQMD does not include any explicit in-medium modifications for vector mesons or effects to describe the restoration of chiral symmetry.
The model allows to study the full
space time evolution of all hadrons, resonances and their decay products in hadron-hadron or nucleus-nucleus collisions.
This permits to explore the emission patterns of resonances
in detail and to gain insight into their origins and decay channels. For previous studies of resonances within this model see \cite{Bleicher:2002dm,Bleicher:2002rx,Bleicher:2003ij,Vogel:2005qr,Vogel:2005pd,Vogel:2007yu,Vogel:2007qg}.
For further details about the UrQMD model the reader is referred to \cite{Bass:1998ca,Bleicher:1999xi}.

Experimentally, the reconstruction of resonances is challenging. One often applied technique is to reconstruct the invariant mass spectrum for single events. Then, an invariant mass distribution of mixed events is generated (here, the particle pairs are uncorrelated by definition). The mixed event distribution is substracted from the invariant mass spectrum of the single (correlated) events. As a result one obtains the mass distributions and yields (after all experimental corrections) of the resonances by fitting the resulting distribution with a suitable function (usually a Breit-Wigner function peaked around the pole mass of the respective resonance).

If the resonance spectral function changes in the hadronic medium this is in principle visible in the difference spectrum between true and mixed events. However, if a daughter particle (re-)scatters before reaching the detector the signal for the experimental reconstruction is blurred or even lost. Especially for strongly interacting decay products this effect can be sizeable. It is therefore difficult to judge whether a deviation from an expected  Breit-Wigner distribution is due to an initial deformation or an increase of the initial width or due to the momentum dependence of the rescattering cross section of the daughter particles.

What makes this analysis even tougher is the fact that the resonances decay over a wide range of densities and therefore only an average value is measured. If this average value is dominated by resonance decays at low density the information from the high density phase is blurred and may offer only a limited view on the high density phase of the heavy ion collision.

UrQMD offers
a different technique for the extraction of resonances which we apply here.
We follow the individual decay products of each
decaying resonance (the daughter particles). If the daughter particles
do not rescatter in the further evolution of the system, the resonance is counted
as ``reconstructable''. The advantage of this method is that it allows
to trace back the origin of each individual resonance to study their spatial and temporal
emission pattern. Because UrQMD follows the space time evolution of all particles 
it is possible to link production and decay point of each individual resonance.
This method also allows to explore the reconstruction efficiency in different decay branches. 

In order to calculate at which density the resonance decays we have to determine the baryonic density.
The baryon density is calculated locally at the position of the resonance in the rest frame of the baryon current (Eckart frame) as $\rho_B=j^0$ with $j^\mu = (\rho_B,\vec{0})$. Details on the calculation of the baryon density
are discussed in \cite{Vogel:2007yu}. In all figures we present the density in units of ground state density, where a value of 0.16 $1/{\rm fm}^3$ is assumed.
In the following we discuss the density dependence of the probability that a resonance can be reconstructed. Naively, one would expect that the higher the densities the more the rescattering effect becomes dominant. Therefore
it is unlikely that a resonance which decays at high density is reconstructable. The view on the low density zone is expected to remain unblurred but is less interesting because it resembles that observed in elementary collisions.

\begin{figure}[hbt]
\vspace*{-0.8cm} 
\epsfig{file=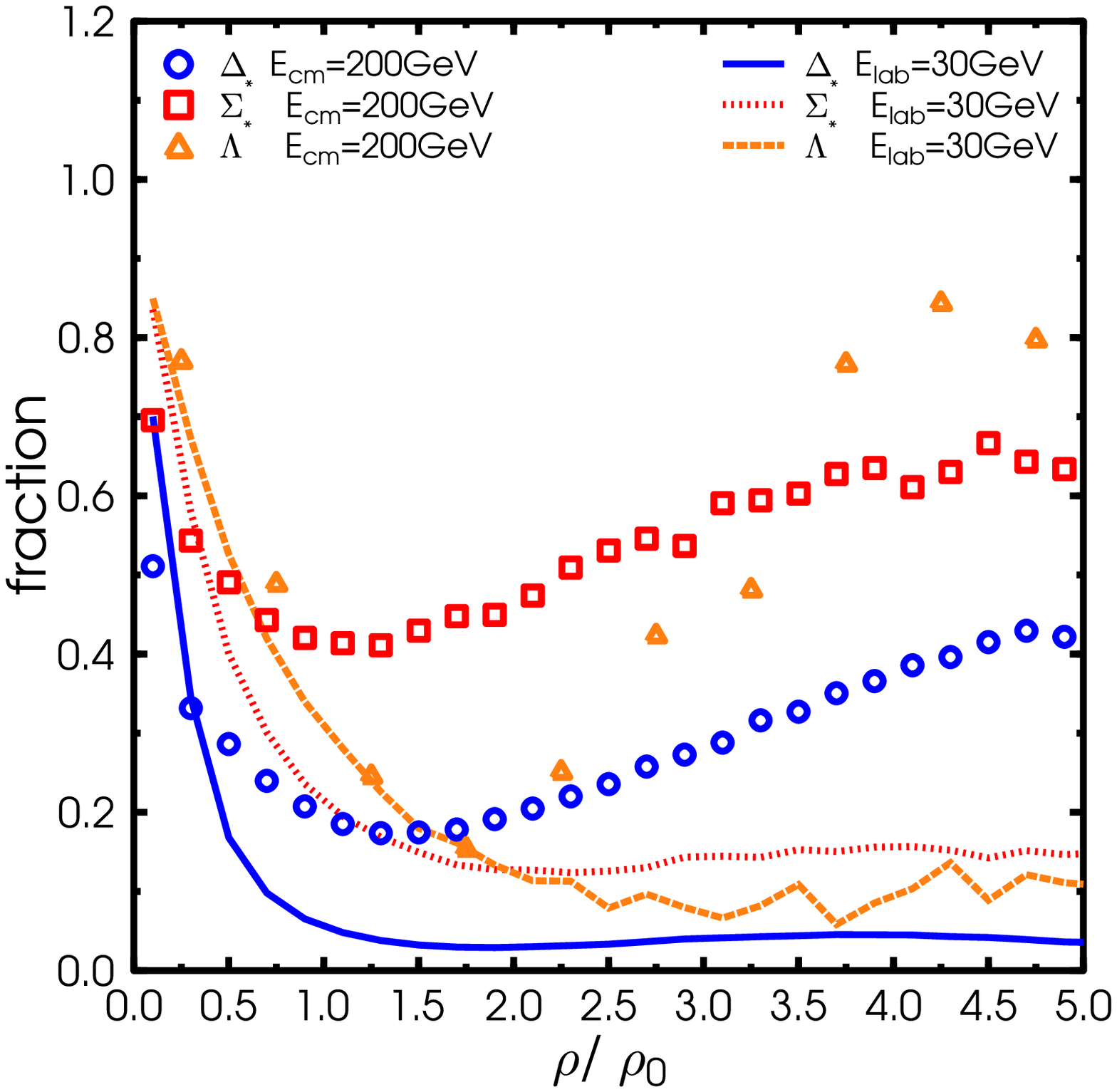,width=7cm}
\epsfig{file=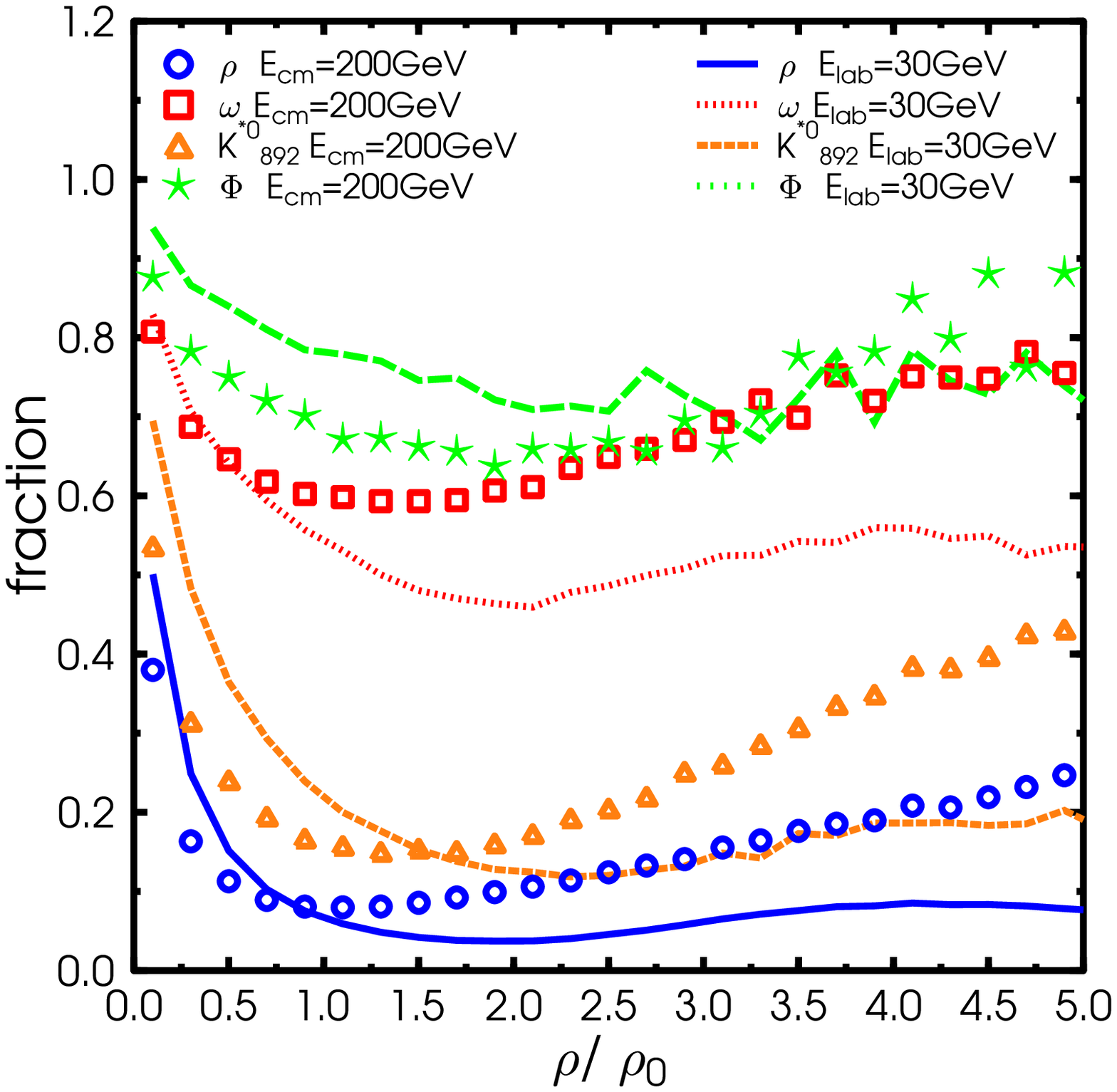,width=7cm}
\caption {(Color online) Fraction of reconstructable baryon resonances (top) and meson resonances (bottom) as a function of baryon density at the point of production.
}
\label{percentage_baryon}
\end{figure}


Depicted in Fig. \ref{percentage_baryon} top (bottom) is the probability that a baryon resonance - shown are $\Delta, \Sigma^*(1385)$ and $\Lambda^*(1520)$ baryon resonances ($\rho, \omega, K^{*0}$ and  $\Phi$ mesons) - which was produced at a certain density can be reconstructed experimentally. One observes a clear peak at very low density and a steady decrease towards higher density. This means that resonances that are produced at rather low density have a high probability to be detected and as the density increases the chance to reconstruct the resonances decreases. This is nothing unexpected. However, this trend stops at roughly 2 $\rho_0$. At higher densities the chance to reconstruct a resonance saturates or even increases slightly again.
This increase, which we discuss later in detail, is caused by resonances which picked up very high transverse momenta and leave the interaction zone quickly. This results in a decay in a region with less hadronic activity and a higher chance to be reconstructed.

Whereas the form of the curves is qualitatively similar for the different hadrons the absolute value of the fraction of reconstructable resonances is rather different. It can be understood in terms of lifetimes of the resonances and in terms of the rescattering cross sections of the decay products.

Due to the large cross section of pions in nuclear matter (usually undergoing $N+\pi \rightarrow \Delta$ or $\pi + \pi \rightarrow \rho$ reactions) the probability to detect a high density $\Delta$ resonance or a $\rho$ meson is rather small compared to the probability to detect a high density $\Phi$ meson, since the $\Phi$ meson itself has a small cross section in nuclear matter and a long lifetime of $\sim$ 40~fm/c and the hadronic decay products (mostly kaons and antikaons) have a smaller cross sections when compared to the pions from the decay of a $\rho$ meson. Similarly,
the long lifetime of the $\Lambda$ increases their possibility to be reconstructed.
As mentioned earlier, the saturation or slight increase of the reconstruction probability as a function of density has its origin in the possibility that resonances with a large $p_T$ can escape quickly from the reaction zone which is rather small initially.

\begin{figure}[t]
\vspace*{-0.8cm} 
\epsfig{file=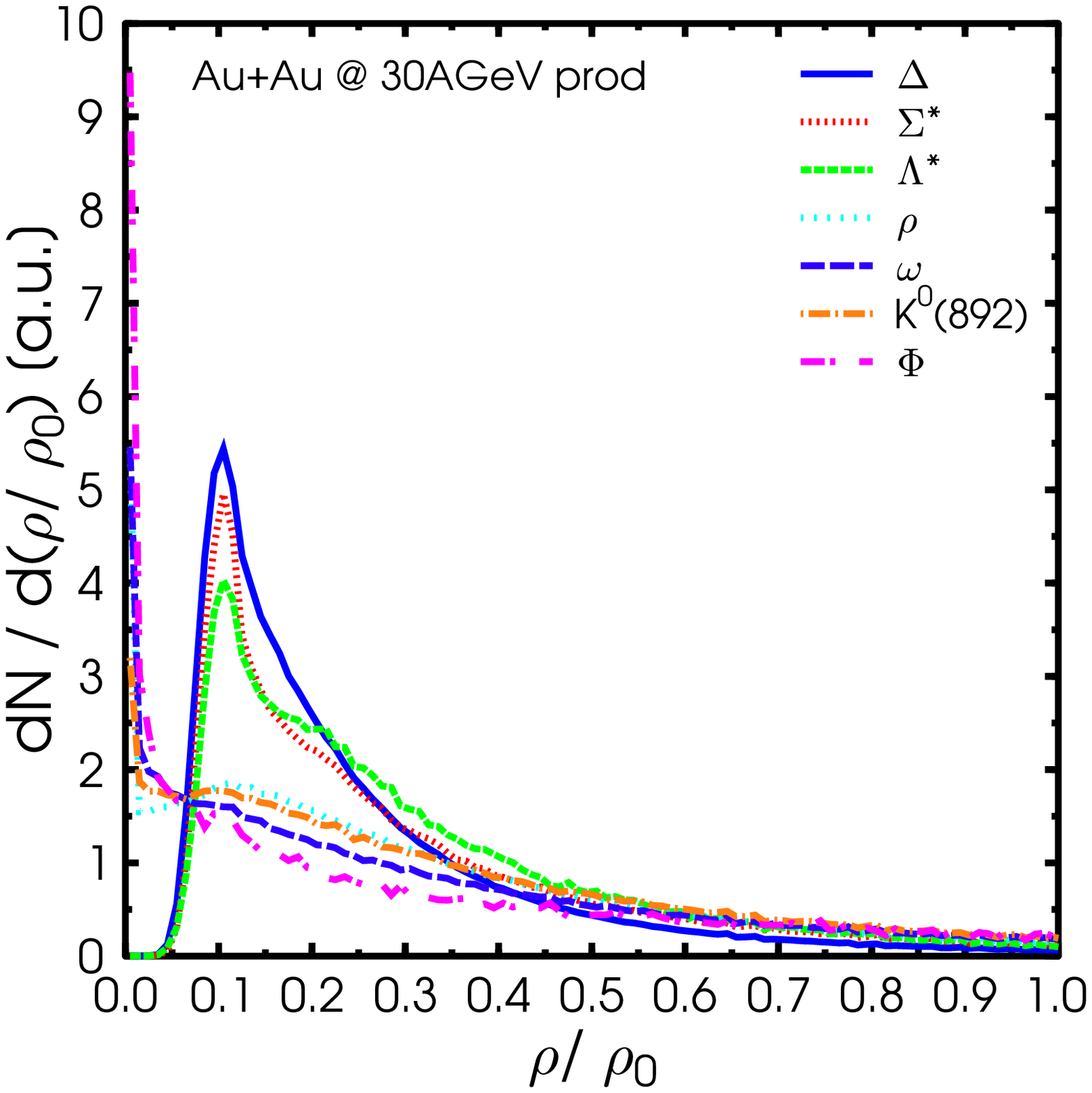,width=7cm}
\epsfig{file=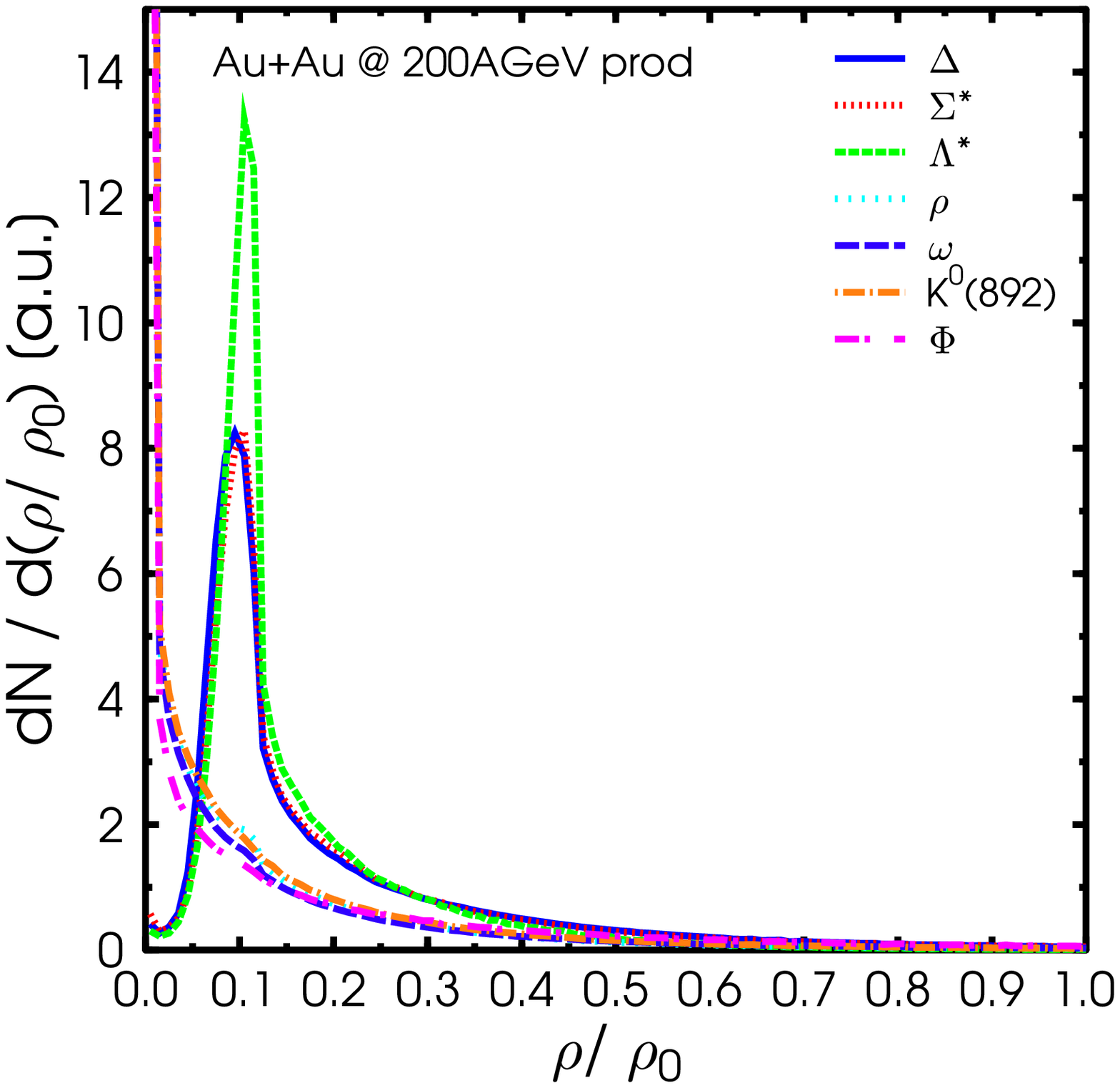,width=7cm}
\caption {(Color online) Probability distribution of baryon density at the production vertex for various reconstructable resonances in central (b $\le$ 3.4~fm) Au+Au collisions at 30~AGeV (top figure) and 200~AGeV (bottom figure) as a function of baryon density. One observes that most resonances which can be reconstructed in the hadronic decay channel originate from low baryon density.}
\label{probability}
\end{figure}

Fig. \ref{probability} shows for various resonances the probability that an experimentally reconstructable resonance has been created at a density $\rho$. The integral over all densities is normalized to unity.
One observes that most of those resonances are produced at very low densities, which is especially true for the mesonic resonances.

Reconstructable baryon resonances stem from slightly higher baryon densities, however most are still produced at rather low densities (with a peak at roughly 0.1 ground state density). So the detection of resonances produced at densities above ground state densities using hadronic decay channels seems not too encouraging. However, as we discuss next, a loophole might exist.\\

Let us illustrate this further with an example which is representative for all investigated particles.

\begin{figure}[hbt]
\vspace*{-0.8cm} 
\epsfig{file=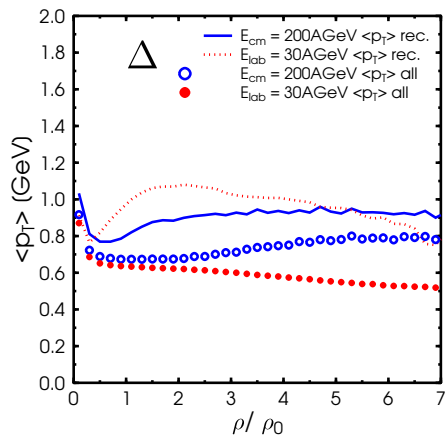,width=7cm} \caption {(Color online) Average transverse momentum of reconstructable (line) or all (symbol) $\Delta$ baryons as a function of baryon density for two different energies.}
\label{delta_pt}
\end{figure}

Fig. \ref{delta_pt} depicts the average transverse momentum of $\Delta$ resonances as a function of baryon density.
Lines show reconstructable resonances, symbols show all decayed resonances. The striking feature is the different average transverse momentum between all resonances and those which are reconstructable. The higher the average transverse momentum, the larger is the chance that the resonance can be reconstructed. The $<p_T>$ of reconstructable $\Delta$ resonances is about  200~MeV higher than for all  $\Delta$ resonances. Resonances with a large $p_T$ can leave the high density zone rather fast and move with a velocity of about $<p_T>/m$ outwards.

 Another interesting feature in Fig. \ref{delta_pt} is the difference between the $\sqrt{s}$=200~AGeV and $E_{lab}$=30~AGeV curves. While the $E_{lab}$=30~AGeV data shows a decrease of $<p_T>$ as a function of the baryon density, the $\sqrt{s}$=200~AGeV data show an increase. At $\sqrt{s}$=200~AGeV the initial collisions (which happen at high baryon density) are  more energetic and give the particles a high transverse momentum, subsequent rescattering
 decreases $p_T$. For the $E_{lab}$=30~AGeV collisions the situation is opposite. Initially the particle $p_T$ is small
 and the rescattering increases the $p_T$ due to transverse expansion.

\begin{figure}[hbt]
\vspace*{-0.8cm} 
\epsfig{file=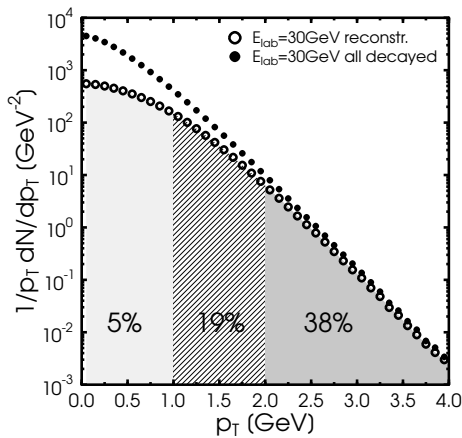,width=7cm} \caption {(Color online)Transverse momentum spectra for all and reconstructable resonances for central (b$\le$3.4~fm) Au+Au collision at 30~AGeV beam energy. Full circles depict the spectrum for all decayed resonances (included in the analysis are $\Delta, \Lambda, \Sigma$ baryons, as well as $\rho, \omega, K^{*0}$ and $\omega$ mesons), open circles for reconstructable resonances. The numbers indicate the percentage of reconstructable resonances stemming from density region with $\rho/\rho_0 > 2$.}
\label{pt_percentages}
\end{figure}

Fig. \ref{pt_percentages} shows the $p_T$ dependence of the reconstruction probability in detail. It shows the transverse momentum spectra for all (full symbols) and reconstructable resonances (open symbols). The numbers stated in the three shaded areas ($p_T < 1~\mathrm{GeV} , 1~\mathrm{GeV} < p_T < 2~\mathrm{GeV} , p_T > 2~\mathrm{GeV}$ ) are the percentages of reconstructable resonances created at a density higher than 2$\rho_0$. One observes that at low transverse momentum the percentage of reconstructable resonances is low and increases when going to higher transverse momenta, i.e. that with increasing $p_T$ the chance to reconstruct a resonance produced at high baryon density increases.

In conclusion, we have discussed that the view on the high density zone may not be as restricted as usually assumed when analyzing hadronic resonances. 

We argued that resonances detected with high transverse momentum are sensitive to higher densities. It will be interesting to explore if the properties of these resonances are different from the bulk emitted at low densities. 
The exploration of high $p_T$ resonances might therefore open a new keyhole at the upcoming CBM experiment at FAIR or the critRHIC program to gain information on the high density zone and to observe eventual changes of resonance properties in the medium.

This work was (financially) supported by the Helmholtz International
Center for FAIR within the framework of the LOEWE program
(Landesoffensive zur Entwicklung Wissenschaftlich-\"Okonomischer Exzellenz)
launched by the State of Hesse. The computational resources have been provided by the Center for the Scientific Computing (CSC) at Frankfurt. This work was supported by GSI, DAAD (PROCOPE) and BMBF.

\end{document}